\documentclass[%
 reprint,
showpacs,preprintnumbers,
 amsmath,amssymb,
 aps,
prc,
superscriptaddress]{revtex4-1}

\usepackage{graphicx}
\usepackage{dcolumn}
\usepackage{bm}
\usepackage[colorlinks=true, pdfstartview=FitV, linkcolor=blue, 
citecolor=blue,urlcolor=navyblue]{hyperref}

\begin{document}

\title{Short-range correlations effects on the deformability of neutron stars} 

\author{Lucas A. Souza$^1$, Mariana Dutra$^1$, C\'esar H. Lenzi$^1$ and Odilon Louren\c{c}o} 

\affiliation{
\mbox{Departamento de F\'isica, Instituto Tecnol\'ogico de Aeron\'autica, DCTA, 
12228-900, S\~ao Jos\'e dos Campos, SP, Brazil}
}
\date{\today}

\begin{abstract}
In the present work, we investigate the effects of short-range correlations (SRC) on the 
dimensionless deformability of the binary neutron system related to the GW170817 event. We 
implemented phenomenological SRC in a relativistic mean-field model in which the bulk parameters, 
namely, incompressibility ($K_0$), effective nucleon mass ratio ($m^*$), symmetry energy ($J$) and 
its slope ($J_0$), are independently controlled. Our results point out that the SRC favor the 
model to pass through the constraints, established by the LIGO/Virgo Collaboration, on the values 
of $\Lambda_{1.4}$ and on the $\Lambda_1\times\Lambda_2$ region. We also found a clear linear 
correlation between $\Lambda_{1.4}$ with $K_0$ and $L_0$ (increasing dependence), and with $m^*$ 
and $J$ (decreasing dependence). Finally, we also obtained compatible numbers for $R_{1.4}$ (model 
with and without SRC) in comparison with recent data from the neutron star interior composition 
explorer mission.
\end{abstract}


\maketitle

\section{Introduction}

A widely approach used to treat many-nucleon systems (nuclei, for example) is based on the 
construction of nuclear interactions, such as the one pion exchange potentials, in which the free 
parameters are adjusted to reproduce, for example, experimental data involved in the simplest bound 
nuclear system used to study the nucleon-nucleon interaction, namely, the deuteron 
(proton-neutron pair). This system has bound state energy of around 
$2$~MeV~\cite{ligacao1,ligacao2}, electric quadrupole momentum of $2.82$~mb~\cite{quadrupolo}, 
magnetic momentum given by $0.86$~mn~\cite{mag}, and it is described by a spin~$1$ and isospin~$0$ 
triplet state. From the knowledge of this interaction, it is possible to treat nuclei by using the
Brueckner-Hartree-Fock~\cite{prc3,bethe} method, for example. From another perspective, it is also 
possible to describe nuclei from the so-called ``nuclear shell model'' (or ``independent particle 
model'')~\cite{shell}, in which it is considered, at first order, that each nucleon moves 
independently and is affected by an average potential due to the remaining nucleons. From the 
solution of the Schr\"odinger equation, the energy levels of the independent nucleon submitted to 
the average potential $V(r)$ are determined. Possible options for $V(r)$ include the Woods-Saxon 
potential~\cite{woods1,woods2}, the harmonic oscillator, or even the finite square well potential. 
 In addition, microscopic systems of many interacting particles, can be described 
by approximation methods in which combinations of high-performance computing techniques allow a 
fundamental understanding of nuclear properties from many-body hamiltonians. Among the main 
many-body models, a more fundamental approach compared to the nuclear shell model is the {\it ab 
initio} method~\cite{abinition,abinition2,abinition3}. It is based on the density functional theory 
(DFT)~\cite{dft,dft2,dft3} in which many-body correlations are combined to the deuteron and 
nucleon-nucleon interactions.

The aforementioned many-body approaches present limitations in the mass number such that a current
challenge is the search for a universal energy density functional including heavy isotopes, 
and able to describe relevant characteristics of finite nuclei and also extended asymmetric 
nucleonic matter.
The shell model is successful, for example, in describing stable nuclei having proton or neutron 
numbers given by 2, 8, 20, 28, 50, 82, 126 (magic numbers). For these cases, the model predicts 
nuclei with filled shells. However, electron-induced quasi-elastic proton knockout 
experiments~\cite{knockout1,knockout2} show that the nucleon-independent particle behavior, in 
which the shell model is based on, occurs to about 70\% of nucleons in valence states. 
Non-independent nucleons correlate in pairs with high relative momentum due to the short-range 
components of the nuclear interaction. These correlations are called short-range correlations 
(SRC)~\cite{nature,hen2017,ye2018,Egiyan2006,Frankfurt1993,Fomin2012,Atti2015,Shneor2007,Tang2003,
Li2019,Schmookler2019,Duer2019,Ryck2019,Chen2017}. Experiments performed at the Thomas Jefferson 
National Accelerator Facility (JLab) concluded that for the $^{12}\rm C$ nucleus, 20\% of the 
nucleons present SRC and, within this set, 90\% of the correlated pairs are neutron-proton~($np$) 
type~\cite{subedi2008}. The remaining pairs are divided into 5\% for each $nn$ and $pp$ pairs. 
The experiments consist of very energetic incident electrons in the $^{12} \rm C$ 
nucleus. In the collision, it is found that a proton is simultaneously removed with another 
correlated nucleon, in this case, a neutron more frequently. The dominance of the correlated $np$ 
pair is also observed in proton removal experiments in heavier nuclei, and even in those with more 
neutrons, as verified in experiments involving $^{27}\rm Al$, $^{56}\rm Fe$, and $^{208}\rm 
Pb$~\cite{orhen} nuclei. The predominance of the specific $np$ pair, or in other words, 
deuteron-like pairs, is justified as a direct consequence of the tensor part of the nuclear 
interaction~\cite{tensor1,tensor2}. 

The implications of SRC, and more specifically, of the predominance in the $np$ pair formation, are 
diverse. For example, in the analysis of neutrino scattering by correlated 
nucleons~\cite{neutrino1,neutrino2}, in the momentum distribution of the quarks that form these 
pair~\cite{quark1,quark2}, and also in many nucleon systems such as nuclear matter. 
Neutron star properties were also shown to be affected by the inclusion of the SRC 
phenomenology as one can verify in Ref.~\cite{cai}.

In this work, we focus on the effects of the SRC in the dimensionless tidal deformability 
($\Lambda$) of the neutron star system related to the GW170817 
event~\cite{ligo17,ligo18,ligo19,tanja19,prc,apj,epja,plb}. Such a recent observation of 
gravitational waves emission, from the binary neutron star merger event, observed on 17 August 2017, 
offered an opportunity to a deep understanding of the stellar matter equation of state (EoS), since 
it provided important constraints that reliable models should satisfy. In Sec.~\ref{rmf}, we present 
the structure of the relativistic mean-field (RMF) model, with SRC included, used to obtain the 
stellar matter EoS and for the calculation of $\Lambda$. In Sec.~\ref{results}, we show the 
influence of the SRC on the deformability of the neutron star binary system. In particular, the 
results point out that the SRC favor the model to reach the GW170817 constraints. Another 
interesting result of our study is the clear linear correlations exhibited between the nuclear 
matter bulk parameters and the deformability. Finally, we present the summary and concluding remarks 
in Sec.~\ref{summary}.

\section{Relativistic mean-field model including short-range correlations}
\label{rmf}

Probes of the short-range correlations can be verified from the analysis of the nucleon momentum 
distribution functions in several nuclei~\cite{ppnp} such as $^{2}\rm H$, $^{4}\rm He$, $^{16}\rm 
O$, $^{40}\rm Ca$, besides $^{12}\rm C$, $^{56}\rm Fe$, and $^{208}\rm Pb$, previously cited. The 
SRC imply a decrease in the occupation of the states below the Fermi level and a partial 
occupation in the states above it. Consequently, the momentum distribution function of nucleons, 
$n(k)$, is different from the step function of a Fermi gas of independent particles, as depicted in 
Fig.~1 of Ref.~\cite{cai}, for instance. The marked area in that figure corresponds to the region 
of the so-called ``high momentum tail''~(HMT), in which $n(k)$ depends on $k$ as $n(k) \sim k^{-4}$.

In Ref.~\cite{cai}, the authors proposed the implementation of the HMT in the equations of state of 
the relativistic mean field model presenting third and fourth order in the scalar field $\sigma$, 
second and fourth order in the vector field $\omega$, and interaction between the $\rho$ and 
$\omega$ mesons. Basically, the effect of the HMT is to change the kinetic momentum integrals of 
the model by replacing the traditional Fermi gas step functions by the new $n(k)$ distributions, in 
this case, given by $n_{n,p}(k_F, y) = \Delta_{n,p}$ for $0<k<k_{F\,{n,p}}$ and $n_{n,p}(k_F, y) = 
C_{n,p}(k_{F\,{n,p}}/k)^4$ for $k_{F\,{n,p}}<k<\phi_{n,p} k_{F\,{n,p}}$. The proton fraction is 
given by $y=\rho_p/\rho$, with $\rho_p$ being the proton density, and $\rho=2k_F^3/(3\pi^2)$ the 
total one. 

In this work, we use the model constructed in Ref.~\cite{cai} that presents $\Delta_{n,p}=1 - 
3C_{n,p}(1-1/\phi_{n,p})$, where $C_p=C_0[1 - C_1(1-2y)]$, $C_n=C_0[1 + C_1(1-2y)]$, 
$\phi_p=\phi_0[1 - \phi_1(1-2y)]$ and $\phi_n=\phi_0[1 + \phi_1(1-2y)]$. Furthermore, we also use 
$C_0=0.161$, $C_1=-0.25$, $\phi_0 = 2.38$ and $\phi_1=-0.56$. These values are determined from 
analysis of $d(e,e',p)$ reactions, medium-energy photonuclear absorptions, $(e, e')$ reactions, and 
data from two-nucleon knockout reactions as described in Ref.~\cite{baoanli}. The energy density and 
pressure of the RMF model with SRC implemented from such an approach are given, respectively, by
\begin{eqnarray} 
\epsilon &=&  \frac{m_{\sigma}^{2} \sigma^{2}}{2} +\frac{A\sigma^{3}}{3} +\frac{B\sigma^{4}}{4} 
-\frac{m_{\omega}^{2} \omega_{0}^{2}}{2} - \frac{Cg_{\omega}^4\omega_{0}^4}{4}
- \frac{m_{\rho}^{2} \bar{\rho}_{0(3)}^{2}}{2} 
\nonumber\\
&+& g_{\omega} \omega_{0} \rho + \frac{g_{\rho}}{2} 
\bar{\rho}_{0(3)} \rho_{3}   -\frac{1}{2} \alpha'_3 g_{\omega}^{2} g_{\rho}^{2} \omega_{0}^{2} 
\bar{\rho}_{0(3)}^{2}
+ \epsilon_{\mathrm{kin}}^{p}+\epsilon_{\mathrm{kin}}^{n}\nonumber\\
\label{eden}
\end{eqnarray}
and
\begin{eqnarray}
p &=&-\frac{m_{\sigma}^{2} \sigma^{2}}{2} - \frac{A\sigma^{3}}{3} - \frac{B\sigma^{4}}{4} 
+ \frac{m_{\omega}^{2} \omega_{0}^{2}}{2} + \frac{Cg_{\omega}^4\omega_0^4}{4}
\nonumber\\
&+& \frac{m_{\rho}^{2} \bar{\rho}_{0(3)}^{2}}{2} + \frac{1}{2} \alpha'_3 g_{\omega}^{2} 
g_{\rho}^{2} \omega_{0}^{2} \bar{\rho}_{0(3)}^{2} + p_{\mathrm{kin}}^{p}+p_{\mathrm{kin}}^{n},
\label{press}
\end{eqnarray}
with the following kinetic contributions
\begin{eqnarray} 
\epsilon_{\text {kin }}^{n,p} &=& \frac{\gamma \Delta_{n,p}}{2\pi^2} \int_0^{{k_{F\,{n,p}}}} 
k^2dk({k^{2}+M^{* 2}})^{1/2}
\nonumber\\
&+& \frac{\gamma C_{n,p}}{2\pi^2} \int_{k_{F\,{n,p}}}^{\phi_{n,p} {k_{F\,{n,p}}}} 
\frac{{k_F}_{n,p}^4}{k^2}\, dk({k^{2}+M^{* 2}})^{1/2},\nonumber \\
\label{ekin}
\end{eqnarray}
and
\begin{eqnarray} 
 p_{\text {kin }}^{n,p} &=&  
\frac{\gamma \Delta_{n,p}}{6\pi^2} \int_0^{k_{F\,{n,p}}}  
\frac{k^4dk}{\left({k^{2}+M^{*2}}\right)^{1/2}} 
\nonumber\\
&+& \frac{\gamma C_{n,p}}{6\pi^2} \int_{k_{F\,{n,p}}}^{\phi_{n,p} {k_{F\,{n,p}}}} 
 \frac{{k_F}_{n,p}^4dk}{\left({k^{2}+M^{*2}}\right)^{1/2}}.
\label{pkin}
\end{eqnarray}
Since the mean-field approximation is taken, $\sigma$, $\omega_0$ (zero component) and 
$\bar{\rho}_{0_{(3)}}$ (isospin space third component) are the expectation values of the mesons 
fields in the expressions above. We also use here that $m_\omega = 782.5$~MeV, $m_\rho = 763$~MeV 
and $m_\sigma=500$~MeV. The effective nucleon mass is defined by $M^*= M_{\mbox{\tiny 
nuc}}-g_\sigma\sigma$, the degeneracy factor is $\gamma=2$ for asymmetric matter, and  
$M_{\mbox{\tiny nuc}}=939$~MeV is the nucleon rest mass. The self-consistency of the model imposes 
to $M^*$ the condition of $M^*- M_{\mbox{\tiny nuc}} + (g_\sigma^2/m_\sigma^2)\rho_s - 
(A/m_\sigma^2)\sigma^2 - (B/m_\sigma^2)\sigma^3 = 0$, with $\rho_s={\rho_s}_p + {\rho_s}_n$ and 
\begin{eqnarray}
{\rho_s}_{n,p} &=& 
\frac{\gamma M^*\Delta_{n,p}}{2\pi^2} \int_0^{k_{F\,{n,p}}}  
\frac{k^2dk}{\left({k^{2}+M^{*2}}\right)^{1/2}} 
\nonumber\\
&+& \frac{\gamma M^*C_{n,p}}{2\pi^2} \int_{k_{F\,{n,p}}}^{\phi_{n,p} {k_{F\,{n,p}}}} 
\frac{{k_F}_{n,p}^4}{k^2}  \frac{dk}{\left({k^{2}+M^{*2}}\right)^{1/2}}.
\nonumber\\
\label{rhos}
\end{eqnarray}
The remaining field equations for $\omega_0$ and $\bar{\rho}_{0(3)}$ are $m_{\omega}^{2} \omega_{0}= 
g_{\omega} \rho-C g_{\omega}\left(g_{\omega} \omega_{0}\right)^{3} -\alpha'_3 g_{\omega}^{2} 
g_{\rho}^{2} \bar{\rho}_{0(3)}^{2} \omega_{0}$ and $m_{\rho}^{2} \bar{\rho}_{0(3)}= g_\rho\rho_3/2- 
\alpha'_3g_{\omega}^{2} g_{\rho}^{2} \bar{\rho}_{0(3)} \omega_{0}^{2}$.

Six free coupling constants, namely, $g_\sigma$, $g_\omega$, $g_\rho$, $A$, $B$, and $\alpha'_3$ are 
determined in order to reproduce six bulk parameters identified as $\rho_0=0.15$~fm$^{-3}$ 
(saturation density), $B_0=-16.0$~MeV (binding energy), $m^*\equiv M^*_0/M_{\mbox{\tiny nuc}}=0.60$ 
(ratio of the effective mass to the nucleon rest mass), $K_0=230$~MeV (incompressibility), 
$J=31.6$~MeV (symmetry energy) and $L_0=58.9$~MeV. Here, one has $B_0=E(\rho_0) - M$, 
$M_0^*=M^*(\rho_0)$, $K_0=9(\partial p/\partial\rho)_{\rho_0}$, $J=\mathcal{S}(\rho_0)$, and 
$L_0=3\rho_0(\partial\mathcal{S}/\partial\rho)_{\rho_0}$, with 
$\mathcal{S}(\rho)=(1/8)(\partial^{2}E/\partial y^2)|_{y=1/2}$ and $E(\rho)=\epsilon/\rho$.

We remind the reader that the values of the bulk parameters related to the ``reference 
model'' applied in Ref.~\cite{cai} and also in our work are based on different theoretical and 
experimental studies. Usually, in nuclear mean-field models, saturation density and binding energy 
are well established closely around the values of $0.15$~fm$^{-3}$ and $-16.0$~MeV, respectively. 
Regarding the symmetry energy and its slope, the authors of Ref.~\cite{plb2013} collected data of 
these quantities from analyses of different terrestrial nuclear experiments and astrophysical 
observations. They included investigations of isospin diffusion, neutron skins, pygmy dipole 
resonances, $\alpha$ and $\beta$ decays, transverse flow, mass-radius relation and torsional crust 
oscillations of neutron stars. From the numbers extracted from these terrestrial laboratory 
measurements and astrophysical observations, they obtained the average values of $J=31.6\pm 
2.66$~MeV and $L_0=58.9\pm 16$~MeV. Another set of data analyzed in Ref.~\cite{jlrange2} provided a 
similar result for these isovector parameters, namely, $J= 31.7 \pm 3.2$ and $L_0=58.7\pm 28.1$. 
Concerning the incompressibility, $K_0=230$~MeV is consistent with the range of 
$220\,\mbox{MeV}\leqslant K_0 \leqslant 260~\mbox{MeV}$ according to the the current consensus on 
this quantity, see Ref.~\cite{k4} for instance. Finally, the value of $0.6$ for the effective mass 
ratio is in agreement with the limits of $0.58\leqslant m^* \leqslant 0.64$~\cite{furns}. In 
Ref.~\cite{furns}, the authors found a strong correlation between $m^*$ and the spin-orbit 
splittings in nuclei, for a particular class of relativistic models. They conclude that 
parametrizations of this model presenting $m^*$ in the mentioned range show spin-orbit splittings in 
agreement with well-established experimental values for $^{16}\rm O$, $^{40}\rm Ca$, and $^{208}\rm 
Pb$ nuclei. Furthermore, the $263$ parametrizations of different kinds of RMF models analyzed in 
Ref.~\cite{rmf} present a range of $0.52\leqslant m^*\leqslant0.80$ for the effective mass ratio. 
The value of $m^*=0.6$ used here is also inside this limit.

The last coupling constant of the model, namely, $C$ is chosen to be $C=0.005$ instead of $C=0.01$ 
from Ref.~\cite{cai}. The former value ensures that the model predicts neutron stars with masses 
around 2 solar masses (see next section).

In order to complete the equations needed to construct the stellar matter, we present the chemical 
potentials for protons and neutrons. They are given by 
\begin{eqnarray}
\mu_{n,p} &=& \frac{\partial \epsilon}{\partial \rho_{n,p}} \nonumber\\
&=& \mu^{n,p}_{\mathrm{kin (SRC)}}+\Delta_{n,p}\mu^{n,p}_\mathrm{kin}+g_{\omega} \omega_{0} 
\pm \frac{g_\rho}{2}\bar{\rho}_{0_{(3)}},
\label{mupn}
\end{eqnarray}
with 
\begin{eqnarray}
 \mu^{n,p}_{\mathrm{kin}}=({k^2_F}_{n,p}+M^{*2})^{1/2}
\end{eqnarray}
and
\begin{align} 
&\mu^{n,p}_{\mathrm{kin\,(SRC)}} = 3 C_{n,p} \left[ \mu^{n,p}_{\mathrm{kin}}
- \frac{\left({\phi_{n,p}^2 {k^2_F}_{n,p} + M^{*2}}\right)^{1/2}}{\phi_{n,p}} \right]
\nonumber\\
&+ {4}C_{n,p} {k_F}_{n,p} \ln\left[\frac{\phi_{n,p} {k_F}_{n,p} + 
\left(\phi_{n,p}^2{k_F^2}_{n,p}+M^{*2}\right)^{1/2} }{ {k_F}_{n,p} + \left(  {k^2_F}_{n,p} + M^{* 
2}\right)^{1/2}}\right].
\nonumber\\
\label{mukin}
\end{align} 
In Eq.~(\ref{mupn}), the sign ($+$) stands for protons and ($-$) for neutrons.

From Eqs.~(\ref{ekin}) and~(\ref{pkin}) one sees that the SRC induce an extra term in the kinetic 
contributions of the model. The scalar density, Eq.~(\ref{rhos}), is also modified in the same 
direction, and the kinetic part of the chemical potentials of the model changes as 
$\mu^{n,p}_\mathrm{kin}\rightarrow\mu^{n,p}_{\mathrm{kin (SRC)}} + 
\Delta_{n,p}\mu^{n,p}_\mathrm{kin}$.

\section{Results}
\label{results}

\subsection{Stellar matter}

In order to determine properties related to the neutron star system, it is needed to take into 
account charge neutrality and $\beta$-equilibrium conditions. We consider stellar matter composed 
by protons, neutrons, electrons and muons, with the last leptons emerging when the electron 
chemical potential exceeds the muon mass, i.e., for $\mu_e=(3\pi^2\rho_e)^{1/3}>m_\mu=105.7$~MeV 
($\rho_e$ is the electron density). Such assumptions lead to the constraints given by  $\mu_n - 
\mu_p = \mu_e=\mu_\mu$ and $\rho_p - \rho_e = \rho_\mu=[(\mu_\mu^2 - m_\mu^2)^{3/2}]/(3\pi^2)$, 
which have to be coupled to the field equations coming from the RMF model. The chemical potential 
and density for the muons are given, respectively, by $\mu_\mu$ and $\rho_\mu$. The total energy 
density and pressure of stellar matter are $\mathcal{E} = \epsilon + \epsilon_e + \epsilon_\mu$ and 
$P = p + p_e + p_\mu$, respectively, with $ \epsilon_l$ and $p_l$ being the energy density and 
pressure of the lepton $l=e,\mu$. Some neutron star properties, such as its mass-radius profile, 
can be found by solving the Tolman-Oppenheimer-Volkoff~(TOV) equations~\cite{tov39,tov39a}.

The spherically symmetric neutron star is composed of a core, described here by the RMF model 
previously presented along with the leptons considered, and a crust, divided into outer and inner 
parts. Due to the restricted knowledge about this specific part of the neutron star 
(there is not a consensus with regard to its exact composition), we decided to treat this region 
with widely used approaches without further assumptions, i.e., without the inclusion of possible 
SRC effects. Moreover, the main purpose of our study is to analyze the effects of SRC on 
$\Lambda_{1.4}$ and, according to Ref.~\cite{poly2}, the more important contribution for this 
quantity comes from the neutron star core EoS, in which we implement the SRC phenomenology. We 
model the outer crust by the Baym-Pethick-Sutherland (BPS) equation of state~\cite{bps} in the 
density region of $6.3\times10^{-12}\,\mbox{fm}^{-3} \leqslant\rho\leqslant 
2.5\times10^{-4}\,\mbox{fm}^{-3}$~\cite{poly2,malik19}. For the inner part, we use the polytropic 
form given by $P(\mathcal{E})=A+B\mathcal{E}^{4/3}$~\cite{cai,poly2,poly1,gogny2} in a range of 
$2.5\times10^{-4}\,\mbox{fm}^{-3} \leqslant\rho\leqslant \rho_t$, where $\rho_t$ is the density 
associated to the core-crust transition found, in our case, by the thermodynamical 
method~\cite{gogny1,cc2,kubis04,gonzalez19}. 

\begin{figure}[!htb] 
\centering
\includegraphics[scale=0.33]{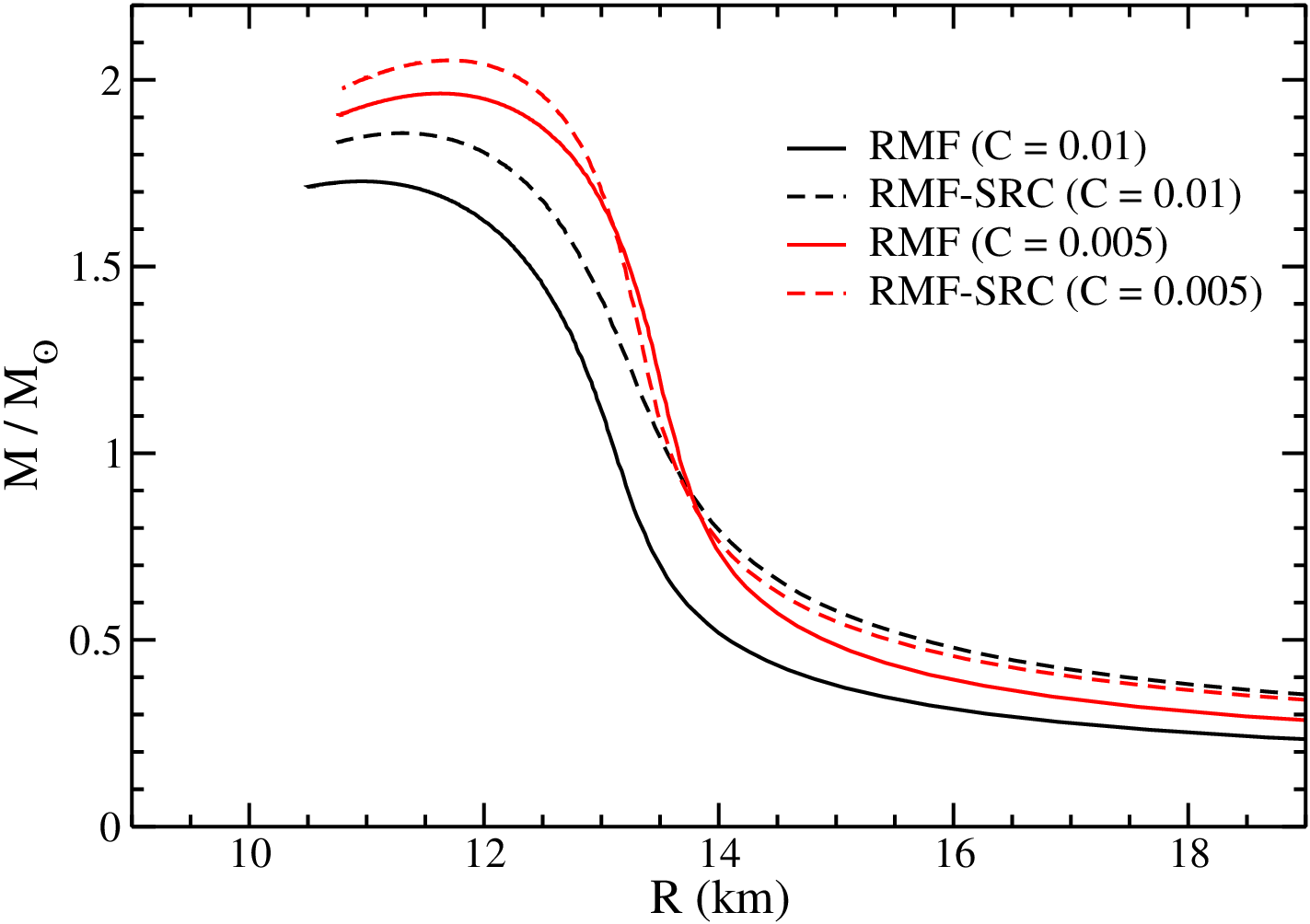}
\caption{Mass-radius diagram for the RMF and RMF-SRC models for $C=0.01$ (black curves) and 
$0.005$ (red curves).} 
\label{mr}
\end{figure}

For both versions of the model, namely, with and without SRC included, we found a maximum neutron 
star mass of $M_{\mbox{\tiny{max}}}=2.05M_\odot$ and $M_{\mbox{\tiny{max}}}=1.96M_\odot$, 
respectively. All of them are compatible with the limits of $(1.928 \pm 
0.017)M_{\odot}$~\cite{fons16,demo10}, $(2.01 \pm 0.04)M_{\odot}$~\cite{anto13}, 
$2.14^{+0.20}_{-0.18} M_\odot$ (95.4\% credible level)~\cite{cromartie} and of 
$2.14^{+0.10}_{-0.09} M_\odot$ (68.3\% credible level)~\cite{cromartie}. It is worth to notice that 
the effect of increasing $M_{\mbox{\tiny{max}}}$ pointed out in Ref.~\cite{cai}, that uses 
$C=0.01$, is also obtained in our model, in which $C=0.005$. In Fig.~\ref{mr}, we 
display the mass-radius profile obtained from the RMF model, with (RMF) and without (RMF-SRC) 
SRC effects, for $C=0.01$ (used in Ref.~\cite{cai}), and for $C=0.005$ (used in 
this work). Notice the difference of $M_{\mbox{\tiny{max}}}$ in both parametrizations due to the 
reduction of the quartic-self coupling strength ($C$ parameter) of the repulsive vector field 
$\omega$.
\begin{widetext}
\textcolor{white}{asdfas}
\begin{figure}[!htb] 
\centering
\includegraphics[scale=0.55]{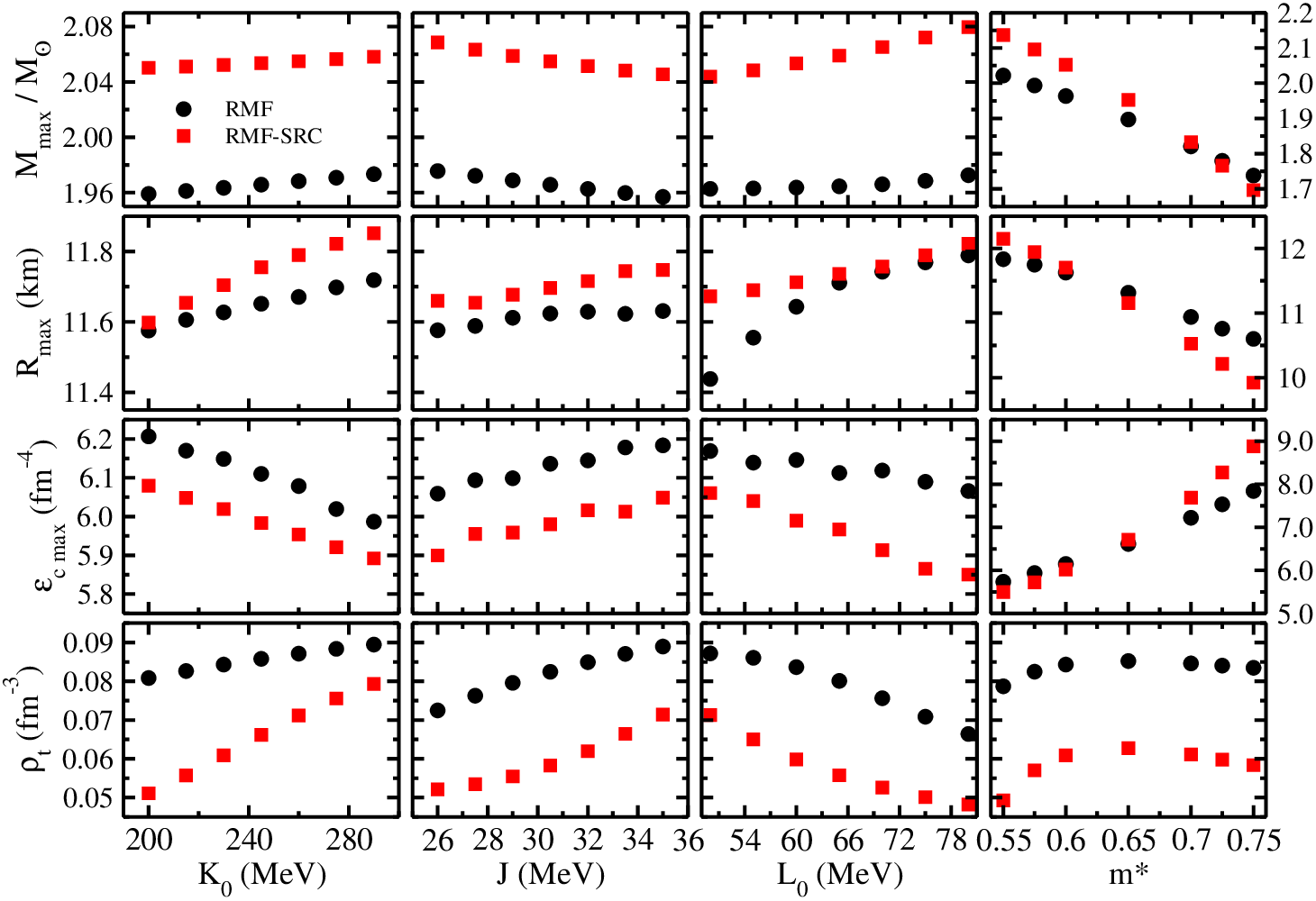}
\caption{Transition density ($\rho_t$) and stellar matter quantities related to the neutron star 
with maximum mass, namely, the maximum mass over $M_\odot$ ($M_{\mbox{\tiny{max}}}/M_\odot$), 
radius ($R_{\mbox{\tiny{max}}}$) and central energy density ($\mathcal{E}_{\mbox{\tiny{c max}}}$), 
as a function of the bulk parameters. } 
\label{stellar}
\end{figure}
\end{widetext}

We are not restricted here to the reference model, namely, the one with the bulk 
parameters given by $\rho_0=0.15$~fm$^{-3}$, $B_0=-16.0$~MeV, $m^*=0.60$, $K_0=230$~MeV, 
$J=31.6$~MeV and $L_0=58.9$~MeV, for $C=0.005$. We also generate different parametrizations by 
changing only one of these quantities while keeping the other ones fixed. We calculate, for RMF and 
RMF-SRC models, some properties related to the maximum mass neutron star for a set of different 
parametrizations. The results are depicted in Fig.~\ref{stellar}, namely, maximum mass 
($M_{\mbox{\tiny{max}}}$), radius ($R_{\mbox{\tiny{max}}}$) and central energy 
density~($\mathcal{E}_{\mbox{\tiny{c max}}}$). The transition densities ($\rho_t$), obtained for 
all the parametrizations studied, are shown as well. 

As one can see all parametrizations, constructed through the variation of the bulk parameters, 
present $M_{\mbox{\tiny{max}}}$ compatible with the observation of $2M_\odot$ millisecond pulsars, 
except for those in which $m^*\geqslant 0.65$ (RMF model). Furthermore, regarding the effect of the 
SRC on the maximum mass, we see that such a phenomenology contributes to increase this quantity, as 
also shown in Fig.~\ref{mr}. In the case of the $m^*$ variation, this increasing is reduced as 
$m^*$ approaches to $0.65$. One can also notice a reduction of $M_{\mbox{\tiny{max}}}$ 
with $m^*$ in both models, feature also registered in Ref.~\cite{tolos}. Another interesting result 
concerning the inclusion of SRC is the reduction of the $\rho_t$ values. This means that the SRC 
enlarge the thermodynamical stable region described by the RMF-SRC model in comparison with the RMF 
one.

\subsection{Deformability calculations (GW170817 event)}

Since the main quantities regarding the stellar matter description are determined, we now focus 
on the deformability calculation related to the neutron star binary system studied in the GW170817 
event~\cite{ligo17,ligo18,ligo19}. For this purpose, we need to obtain the dimensionless tidal 
deformability, written in terms of the (second) tidal Love number $k_2$ as $\Lambda = 
2k_2/(3C^5)$, with $C=M/R$ ($M$ and $R$ are the mass and radius, respectively, of the neutron 
star). $k_2$ is evaluated through the following expression,
\begin{eqnarray}
k_2 &=&\frac{8C^5}{5}(1-2C)^2[2+2C(y_R-1)-y_R]\nonumber\\
&\times&\Big\{2C [6-3y_R+3C(5y_R-8)] \nonumber\\
&+& 4C^3[13-11y_R+C(3y_R-2) + 2C^2(1+y_R)]\nonumber\\
&+& 3(1-2C)^2[2-y_R+2C(y_R-1)]{\rm ln}(1-2C)\Big\}^{-1},\quad
\label{k2}
\end{eqnarray}
with $y_R\equiv y(R)$. The function $y(r)$ is obtained through the solution of a differential 
equation solved as part of a coupled system of equations containing the TOV 
ones~\cite{poly2,tanj10,new,hind08,damour,tayl09}. 

\begin{figure}[!htb]
\centering
\includegraphics[scale=0.35]{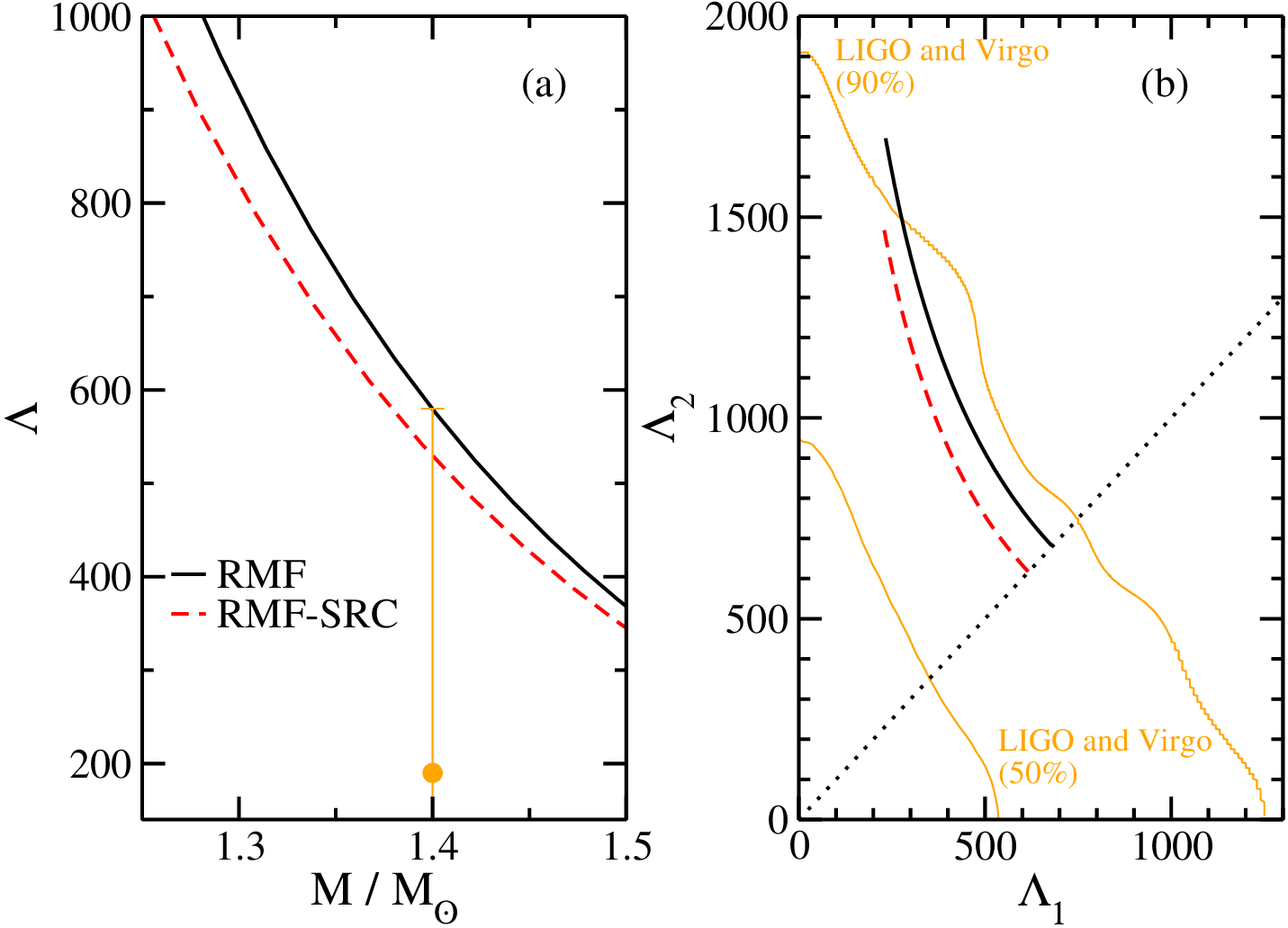}
\caption{Results from the model with (\mbox{RMF-SRC}) and without (RMF) SRC effects. (a) $\Lambda$ 
as a function of $M$. Full circle: result of $\Lambda_{1.4}=190^{+390}_{-120}$ from 
Ref.~\cite{ligo18}. (b) Dimensionless tidal deformabilities for the case of high-mass ($\Lambda_1$) 
and low-mass ($\Lambda_2$) components of the GW170817 event. The confidence lines (90\% and 50\%) 
are also taken from Ref.~\cite{ligo18}.} 
\label{rmf-src}
\end{figure}

In Fig.~\ref{rmf-src}, we show the results of $\Lambda$ as a function of the neutron star mass and 
the dimensionless deformabilities related to the binary system of the GW170817 event.
For the sake of comparison, we display results for the model with and without short-range 
correlations included. From the figure, it is clear that SRC favor the model to reach the GW170817 
data, since $\Lambda_{1.4}$ decreases in comparison with the model without the 
effects, see panel (a), and the $\Lambda_2\times\Lambda_1$ curve moves to the direction of the 
inner region of the LIGO and Virgo Collaboration (LVC) data, as can be seen in panel (b). In the 
calculations of Fig.~\ref{rmf-src}{\color{blue}b}, we use the range of $1.365\leqslant m_1/M_\odot 
\leqslant 1.60$~\cite{ligo17} for the neutrons star with mass $m_1$ from the binary system, and the 
corresponding mass of the companion star, obtained through $[(m_1m_2)^{3/5}]/[(m_1+m_2)^{1/5}] = 
1.188M_\odot$~\cite{ligo17}.

\begin{figure}[!htb]
\centering
\includegraphics[scale=0.355]{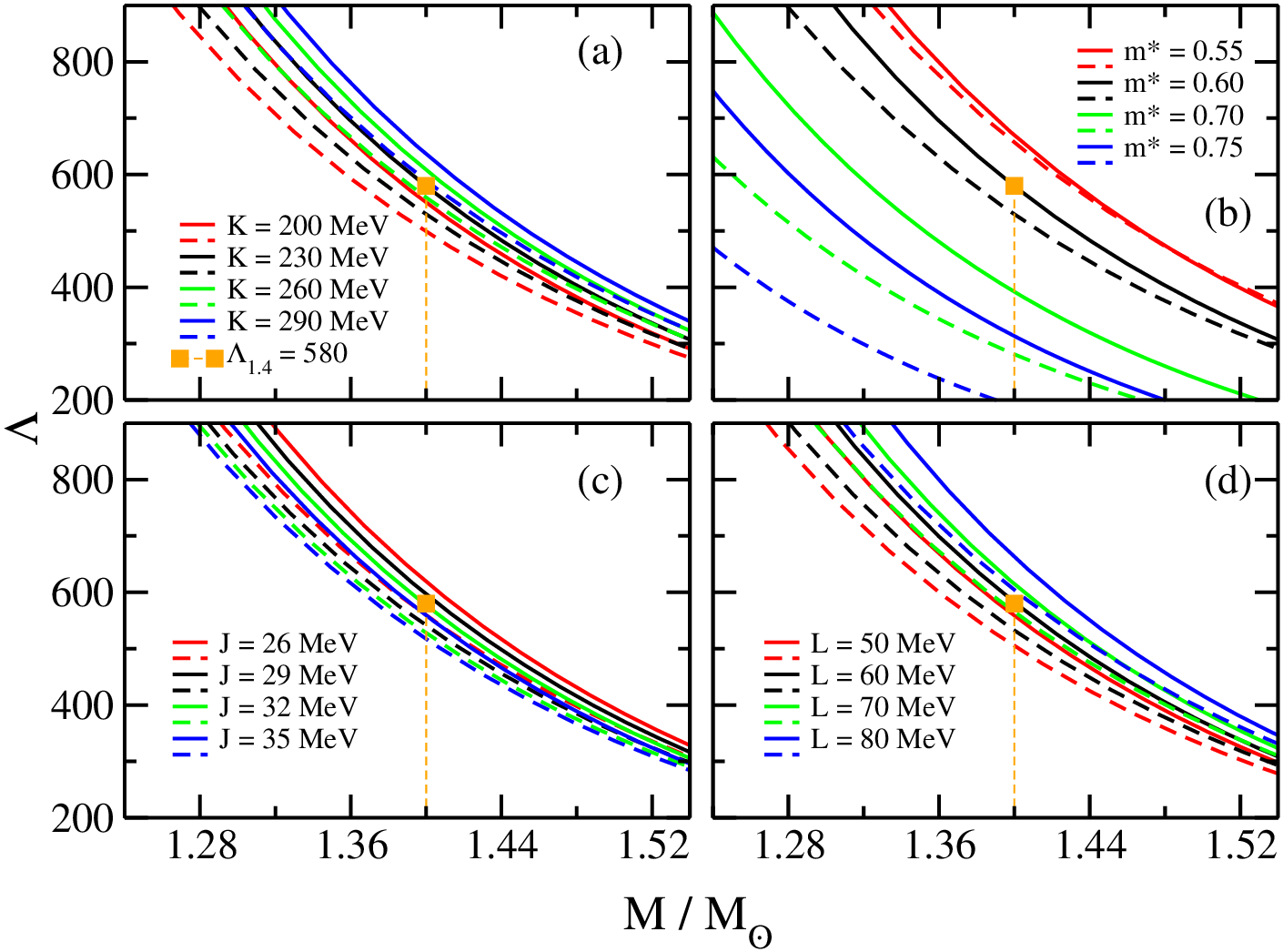}
\caption{$\Lambda$ as a function of the neutron star mass for different parametrizations of the RMF 
model with (dashed lines) and without (full lines) SRC included. Orange square: upper limit of 
$\Lambda_{1.4}=190^{+390}_{-120}$.} 
\label{lm}
\end{figure}

\begin{figure}[!htb]
\centering
\includegraphics[scale=0.342]{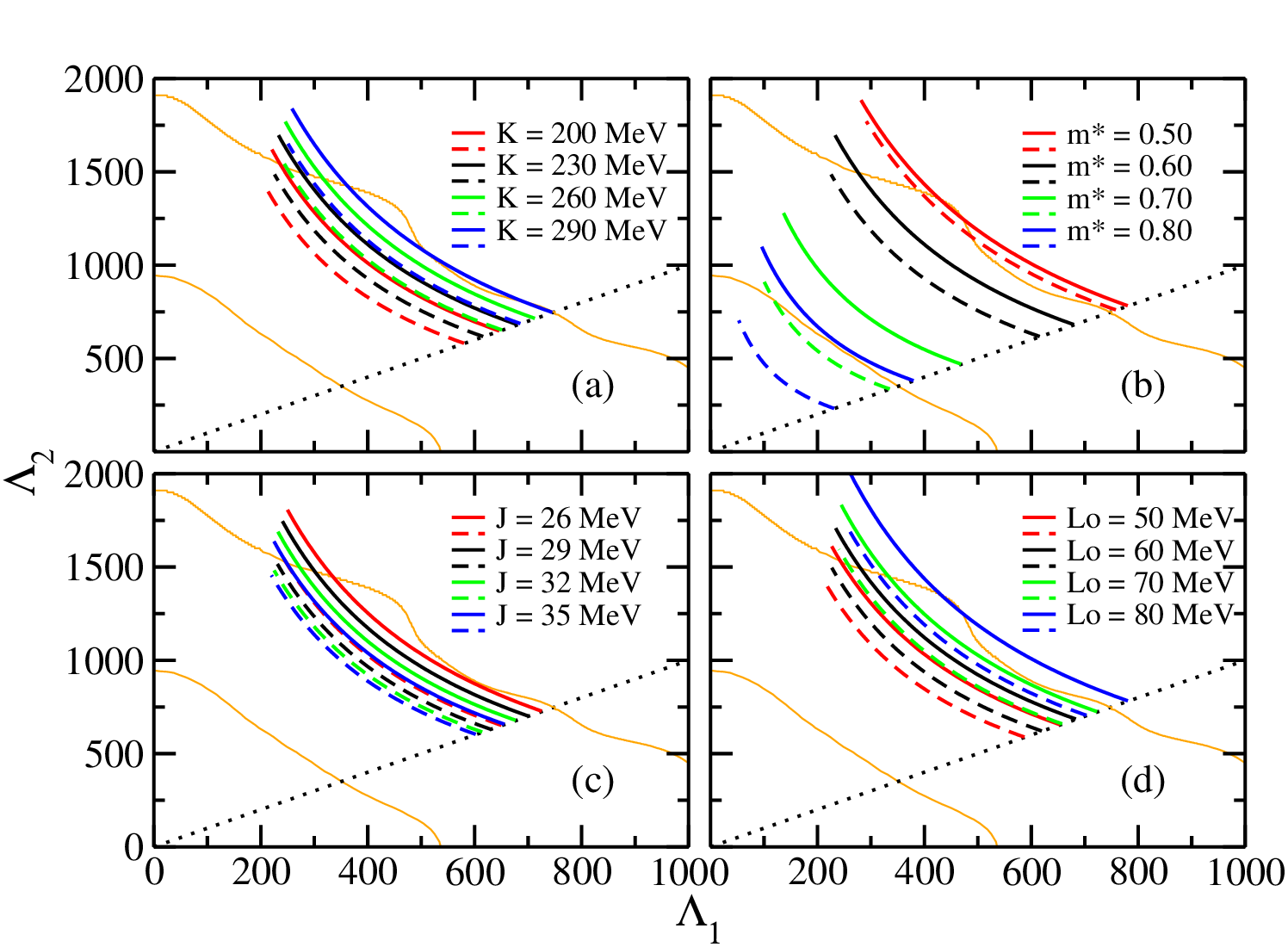}
\caption{$\Lambda_1\times\Lambda_2$ for different parametrizations of the RMF model with (dashed 
lines) and without (full lines) SRC included. Orange lines: confidence lines of 90\% and 50\% from 
Ref.~\cite{ligo18}.} 
\label{l1l2}
\end{figure}

We also verified whether the SRC effects are restricted or not to the reference model used in this 
work, by generating different parametrizations obtained from the independent variation 
of each bulk parameter. In that way, we ensure the particular effect of the specific 
quantity we are changing. In Figs.~\ref{lm} and~\ref{l1l2} we display the results for such new 
parametrizations.

For example, in Figs.~\ref{lm}{\color{blue}a} and~\ref{l1l2}{\color{blue}a} we generate four 
parametrizations, each one with $\rho_0=0.15$~fm$^{-3}$, $B_0=-16.0$~MeV, $m^*=0.60$, $J=31.6$~MeV 
and $L_0=58.9$~MeV fixed, but changing $K_0$ as indicated in the panels. For each particular 
parametrization, we tested the effect of the SRC. The same procedure is performed for the other 
isoscalar and isovector bulk parameters in the remaining panels. The results show the same behavior 
presented in Fig.~\ref{rmf-src}, i.e., the SRC affects the dimensionless deformabilities, namely, 
$\Lambda_{1.4}$ and the $\Lambda_1-\Lambda_2$ pair, always in the direction of the LVC observational 
data. The smaller effect of the SRC inclusion is observed for the parametrization for which 
$m^*=0.55$. Nevertheless, it is also observed that the SRC effects are more pronounced for 
parametrizations with higher values of $m^*$. Notice that the differences between the models 
with and without SRC are higher for parametrizations with $m^*\geqslant 0.60$. Since $M^*= 
M_{\mbox{\tiny nuc}}-g_\sigma\sigma$, it is possible to say that the attractive interaction, 
represented by the scalar $\sigma$ filed, plays the major role regarding the effects produced by 
the inclusion of the SRC in the model. Such a feature is verified in the deformability calculations.

\begin{figure}[!htb]
\centering
\includegraphics[scale=0.355]{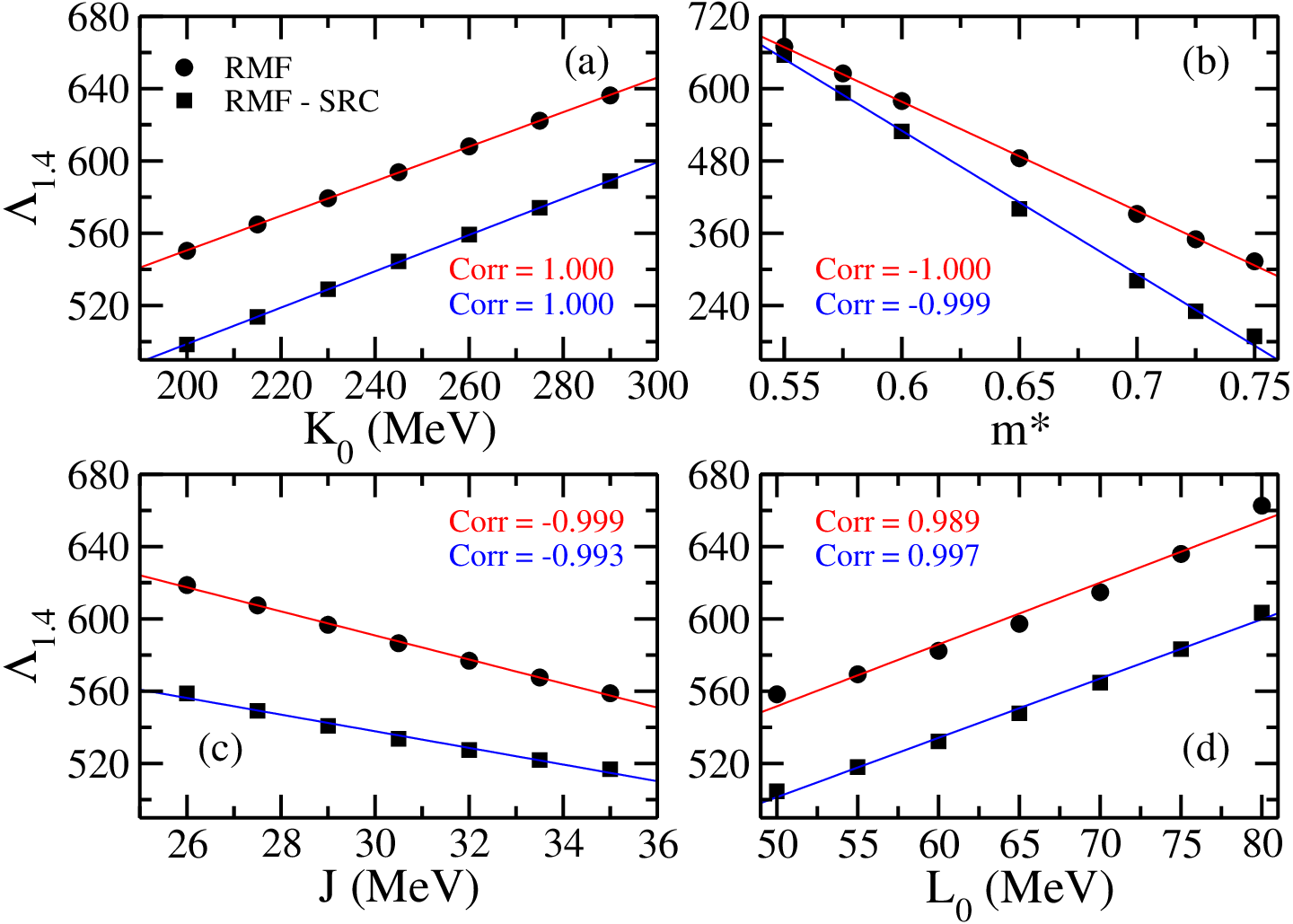}
\caption{$\Lambda_{1.4}$ as a function of the bulk parameters: (a)~$K_0$, (b)~$m^*$, (c)~$J$ and 
(d)~$L_0$. Results for the model with (squares) and without (circles) short-range correlations 
included. Full lines: fitting curves.} 
\label{correlations}
\end{figure}

Since we are able to generate different parametrizations of the RMF model by changing independently 
its bulk parameters, we performed an investigation on the impact of these quantities in the 
dimensionless deformability related to the canonical star ($M=1.4M_\odot$), 
namely,~$\Lambda_{1.4}$. The results are depicted in Fig.~\ref{correlations}. For all 
parametrizations, $\rho_0=0.15$~fm$^{-3}$ and $B_0=-16.0$~MeV are fixed.
From the figure, one can notice a clear linear correlation between $\Lambda_{1.4}$ and the bulk 
parameters, with correlation coefficients around $1$. Such relationships are preserved even when 
the SRC are included in the model. For this case, it is observed that this phenomenology favors the 
model to reach the limits of $\Lambda_{1.4}=190^{+390}_{-120}$ from the LVC, corroborating the 
findings exhibited in Fig.~\ref{lm}. Furthermore, it is also clear that $\Lambda_{1.4}$ is more 
sensitive to variations of $m^*$, as we discussed above. Such a pattern is confirmed in 
Fig.~\ref{correlations}{\color{blue}b}, with a linear dependence clearly established. We also notice 
that $\Lambda_{1.4}$ is an increasing function of $K_0$ or $L_0$, and decreases with $J$ or~$m^*$. 
This feature is also presented for the parametrizations of the \mbox{RMF-SRC} model. Regarding the 
$L_0$ dependence of $\Lambda_{1.4}$, it is worth noticing that such a pattern (increasing of 
$\Lambda_{1.4}$ with $L_0$) was also observed in Ref.~\cite{zhang}, in which the authors used the 
empirical parabolic law for the energy per particle as a function of the density and the 
isospin asymmetry $\delta = 1 - 2y$. Their findings and ours can be indicative that the 
$\Lambda_{1.4}\times L_0$ function may follows an increasing behavior, and as we have shown, the SRC 
do not break this pattern. 

Other interesting results, obtained from the analysis of Fig.~\ref{correlations}, are the ranges of 
the bulk parameters extracted from the relationships of these quantities with the limits of 
$\Lambda_{1.4}=190^{+390}_{-120}$. Since $K_0$ and $L_0$ can be negatives, from the linear 
fitting curves, we focus on their maximum values, related to the upper limit of $\Lambda_{1.4}$. 
They are given by $230$~MeV and $58$~MeV, respectively, for the RMF model. On the other hand, when 
SRC are included, these numbers change to $K_0=280$~MeV and $L_0=74$~MeV. For the incompressibility, 
it is found some overlap of these ranges (with and without SRC) and the current consensus of 
$220\,\mbox{MeV}\leqslant K_0 \leqslant 260~\mbox{MeV}$~\citep{k4}. Concerning $L_0$, it is also 
found an intersection with the range of $25\,\mbox{MeV}\leqslant L_0 \leqslant 
115~\mbox{MeV}$~\cite{rmf}, or even the more stringent ones given by $L_0=58.9\pm16$~MeV~\cite{cai} 
and $L_0=58.7\pm 28.1$~MeV~\cite{jlrange2}, for instance. For both quantities, it is verified that 
the SRC enlarge the overlaps between the values of $K_0$ and $L_0$ estimated from the LVC data and 
the ones found by other predictions. Concerning the limits of $J$ and $m^*$, the fitting curves do 
not produce negative values for these quantities. For the symmetry energy, the limits found are 
quite large, namely, $31.6\,\mbox{MeV}\leqslant J \leqslant 108~\mbox{MeV}$ (RMF) and 
$20.8\,\mbox{MeV}\leqslant J \leqslant 131~\mbox{MeV}$ (\mbox{RMF-SRC}), in comparison with $J= 
31.6 \pm 2.66$~MeV~\cite{cai}, $J= 31.7 \pm 3.2$~MeV~\cite{jlrange2} and $25\,\mbox{MeV}\leqslant 
J \leqslant 35~\mbox{MeV}$~\cite{rmf}. For the effective mass, the ranges are given by 
$0.60\leqslant m^* \leqslant 0.88$ and $0.58\leqslant m^* \leqslant 0.79$, respectively, for the 
RMF model and the \mbox{RMF-SRC} one. Unlike the ranges of $K_0$, $L_0$ and $J$, the SRC reduce the 
range of $m^*$ obtained through the association with $\Lambda_{1.4}=190^{+390}_{-120}$.

As a remark, we emphasize to the reader that the effect of the bulk parameters on 
$\Lambda_{1.4}$ displayed in Fig.~\ref{correlations} (increasing or decreasing, at least) is not 
universal concerning all quantities, namely, $K_0$, $m^*$, $J$ and $L_0$. In Ref.~\cite{apj}, for 
instance, it was observed that $\Lambda_{1.4}$ increases as a function of $K_0$ in a density 
dependent van der Waals model (nucleon-nucleon interactions parametrized as a function of the 
density). This is the same pattern observed in the RMF/\mbox{RMF-SRC} models. However, in that model 
$\Lambda_{1.4}$ increases as $J$ increases, showing the opposite behavior in comparison with 
Fig.~\ref{correlations}{\color{blue}c}. Furthermore, this opposite dependence is also presented in 
the RMF/\mbox{RMF-SRC} models in which $C=\alpha'_3=0$ in Eqs.~(\ref{eden}) and~(\ref{press}) (not 
shown). For these models, there is no restriction on the symmetry energy slope as in the models 
studied here. With regard to the $m^*$ dependence of $\Lambda_{1.4}$, the decreasing of the 
latter as a function of the former is also observed in Ref.~\cite{tolos}, where the authors 
investigate a RMF model with $C=0$. Lastly, the effect of $L_0$ is also shown to be of increasing 
in $\Lambda_{1.4}$ for nonrelativistic Gogny and MDI models studied in Ref.~\cite{plb}. In this 
reference, the authors also found a clear linear correlation for $\Lambda_{1.4}\times L_0$. 
The increasing of $\Lambda_{1.4}$ due to the increasing of $L_0$ was also verified in 
Refs.~\cite{jpg,zhang}.

Finally, by restricting our calculations to the ranges for $K_0$, $L_0$, $J$ and $m^*$ given by 
the circles and squares presented in Fig.~\ref{correlations}, we obtained the following results for 
the radius of the neutron star with $M\sim 1.4M_\odot$: $12.45\,\mbox{km}\leqslant R_{1.4} 
\leqslant 13.71~\mbox{km}$ (RMF) and $11.51\,\mbox{km}\leqslant R_{1.4} \leqslant 
13.61~\mbox{km}$ (\mbox{RMF-SRC}). Such limits are in agreement with the recent findings related to 
the millisecond pulsar PSR J0030+0451, namely, $R_{1.4}=13.89^{+1.22}_{-1.39}$~km~\cite{nicer1} 
and $R_{1.4}=13.02^{+1.24}_{-1.06}$~km~\cite{nicer2}, determined from the data coming from the 
NASA's Neutron Star Interior Composition Explorer (NICER) mission.

\section{Summary and concluding remarks}
\label{summary}

In this work we analyzed the effects of the short-range correlations (SRC) on the dimensionless 
deformability related to the binary neutron star system of the GW170817 event. For the RMF model 
used in this work, in which $\rho_0=0.15$~fm$^{-3}$, $B_0=-16.0$~MeV, $m^*=0.60$, 
$K_0=230$~MeV, $J=31.6$~MeV and $L_0=58.9$~MeV, we verified that the inclusion of the SRC favor the 
model to reach the constraint of $\Lambda_{1.4}=190^{+390}_{-120}$ (regarding the neutron star of 
$M=1.4M_\odot$) and that one observed in the $\Lambda_1\times\Lambda_2$ region, as we shown in 
Fig.~\ref{rmf-src}. This feature is not restricted to this particular model. We verified that the 
impact of the SRC is the same even for different parametrizations (different bulk parameters), as 
exhibited in Figs.~\ref{lm} and~\ref{l1l2}. The SRC are more pronounced with respect to variations 
of $m^*$, which shows a more important role of the attractive interaction represented by the scalar 
field $\sigma$.

We also analyzed that $\Lambda_{1.4}$ is strongly correlated with the isoscalar quantities $K_0$ and 
$m^*$, and the isovector ones $J$ and $L_0$. The relationships remain the same, i.e., a linear 
dependence, even when SRC are included, according to the findings pointed out in 
Fig.~\ref{correlations}. It was also verified that the ranges for the bulk parameters, associated 
with $\Lambda_{1.4}=190^{+390}_{-120}$, present some overlap with other constraints on $K_0$, $J$ and 
$L_0$ obtained from different predictions. Finally, the calculations with the RMF and 
\mbox{RMF-SRC} models pointed out to compatible numbers for $R_{1.4}$ in comparison with the data 
obtained from the Neutron Star Interior Composition Explorer (NICER) mission.

\acknowledgments
This work is a part of the project INCT-FNA Proc. No. 464898/2014-5, partially supported by 
Conselho Nacional de Desenvolvimento Cient\'ifico e Tecnol\'ogico (CNPq) under grants 310242/2017-7 
and 406958/2018-1 (O.L.), 433369/2018-3 (M.D.), and by Funda\c{c}\~ao de Amparo \`a Pesquisa do 
Estado de S\~ao Paulo (FAPESP) under the thematic projects 2013/26258-4 (O.L.), 2019/07767-1 
(L.A.S.) and 2017/05660-0 (O.L., M.D.). L.A.S. thanks the support by INCT-FNA Proc. No. 
88887.464764/2019-00.

\end{document}